# Dimer rattling mode induced low thermal conductivity in an excellent acoustic conductor

Ji Qi[1,2], Baojuan Dong[1,12,13], Zhe Zhang[1,2], Zhao Zhang[1,2], Yanna Chen[3], Qiang Zhang[4], Sergey Danilkin[5], Xi Chen[6,7], Liangwei Fu[8], Xiaoming Jiang[9], Guozhi Chai[10], Satoshi Hiroi[3], Koji Ohara[11], Zongteng Zhang[1,2], Weijun Ren[1], Teng Yang[1,2], Jianshi Zhou[6], Sakata Osami[3], Jiaqing He[8], Dehong Yu[5*], Bing Li[1,2*], Zhidong Zhang[1,2]

[1]Shenyang National Laboratory for Materials Science, Institute of Metal Research, Chinese Academy of Sciences, 72 Wenhua Road, 110016 Shenyang, China.
[2]School of Materials Science and Engineering, University of Science and Technology of China, Shenyang 110016, China.
[3]Synchrontron X-ray station at SPring-8, Research Network and Facility Services Division, National Institute for Materials Science (NIMS), 1-1-1 Kouto, Sayo-Cho, Sayo-gun, Hyogo, 679-5148, Japan.
[4]Spallation Neutron Source, Oak Ridge National Laboratory, Oak Ridge, Tennessee 37831, USA.
[5]Australian Nuclear Science and Technology Organisation, Locked Bag 2001, Kirrawee DC NSW 2232, Australia.
[6]Department of Mechanical Engineering, University of Texas at Austin, Austin, TX 78712, USA.
[7]Department of Electrical and Computer Engineering, University of California, Riverside, CA 92521, USA.
[8]Department of Physics, South University of Science and Technology of China, Shenzhen 518005, China.
[9]State Key Laboratory of Structural Chemistry, Fujian Institute of Research on the Structure of Matter, Chinese Academy of Sciences, Fuzhou, Fujian 350002, China.
[10]Key Lab for Magnetism and Magnetic Materials of the Ministry of Education, Lanzhou University, Lanzhou 730000, China.
[11]SPring-8, Diffraction and Scattering Division, Japan Synchrotron Radiation Research Institute, 1-1-1 Kouto, Sayo-Cho, Sayo-gun, Hyogo, 679-5198, Japan.
[12]State Key Laboratory of Quantum Optics and Quantum Optics Devices, Institute of Opto-Electronics, Shanxi University, Taiyuan, 030006, China.
[13]Collaborative Innovation Center of Extreme Optics, Shanxi University, Taiyuan 030006, China.

*Corresponding email: bingli@imr.ac.cn or dyu@ansto.gov.au

**A solid with larger sound speeds exhibits higher lattice thermal conductivity ($\kappa_{lat}$) [1-3]. Diamond[4] is a prominent instance where its mean sound speed is 14400 m s$^{-1}$ and $\kappa_{lat}$ is 2300 W m$^{-1}$ K$^{-1}$. Here, we report an extreme exception that CuP$_2$ has quite large mean sound speeds of 4155 m s$^{-1}$, comparable to GaAs[5,6], but the single crystals show a very low lattice thermal conductivity of about 4 W m$^{-1}$ K$^{-1}$ at room temperature, one order of magnitude smaller than GaAs. To understand such a puzzling thermal transport behavior, we have thoroughly investigated the atomic structure and lattice dynamics by combining neutron scattering techniques with first-principles simulations. Cu atoms form dimers sandwiched in between the layered P atomic networks and the dimers vibrate as a rattling mode with frequency around 11 meV. This mode is manifested to be remarkably anharmonic and strongly scatters acoustic phonons to achieve the low $\kappa_{lat}$. Such a dimer rattling behavior in layered structures might offer an unprecedented strategy for suppressing thermal conduction without involving atomic disorder.**

Thermal conduction is one of the most fundamental physical properties of materials[7]. Materials with low thermal conductivity are desirable for a great variety of applications such as thermal insulation[8], phase transition memory devices[9] and thermoelectric energy conversion[10]. In electrically insulating nonmagnetic systems, phonons are the major heat carrier and the lattice thermal conductivity is proportional to the product of square of sound speeds and phonon lifetimes[3]. To account for relationship between thermal conductivity and sound speeds, we summarize a variety of disorder-free compounds in **Fig. 1a**. In the logarithmic scale, most of compounds are distributed around an empirical straight line. For a given thermal conductivity value the sound speeds are the dominating factor for materials above the line while the lattice anharmonicity governs thermal transports of materials below the line. To decrease the thermal conductivity, a general approach is to suppress the phonon lifetimes through enhancing phonon-phonon and phonon-disorder scattering. Phonon-phonon scattering, also known as phonon anharmonic interactions, dominates the thermal conductivity of disorder-free systems like PbTe[11] and SnSe[12]. Among these examples, there is a special case called rattling, *i.e.*, a localized vibrational mode scatters acoustic phonons with indication of anti-crossing points in the phonon dispersions[13]. This idea has been broadly applied to rationalize the low thermal conductivity in phonon-glasses thermoelectric materials like filled skutterudites and clathrates[14-16]. In addition, spatially hierarchical atomic disorder has also been widely used, including solid solution of alloy[17], nanostructures[18,19], liquid-like disorder[20,21] and so forth, to lower the thermal conductivity through phonon-disorder scattering.

Recently, we have found that $CuP_2$ exhibits a very low thermal conductivity at room temperature. The temperature dependencies of the lattice thermal conductivity ($\kappa_{lat}$) of the single crystals along *bc* plane and out of *bc* plane are shown in **Fig. 1b** in the temperature region from 50 to 300 K, respectively. The $\kappa_{lat}$ values in both directions increase as temperature decreases. At 300 K, $\kappa_{lat}$ along the *bc* plane is 4.66 W m$^{-1}$ K$^{-1}$ while 3.57 W m$^{-1}$ K$^{-1}$ out of the *bc* plane. $\kappa_{lat}$ for a polycrystalline sample is shown in Supplementary Fig. 1. At room temperature, it is about 0.62 W m$^{-1}$ K$^{-1}$. This value is much reduced compared to the single crystals, which might be attributed to the grain boundaries and defects scattering[22]. In order to understand the mechanism responsible for such a low thermal conductivity in $CuP_2$, we conducted a systematical study of the atomic structure and lattice dynamics for this material.

This compound crystallizes in a relatively simple monoclinic structure with the space group of *P*2$_1$/*c* (ref. 23). As shown in **Fig. 1c**, the structure is characteristic of a layered configuration where Cu layers (spacing 1.59 Å) and P network (spacing 3.85 Å) repeat alternatively along the *a* axis. The Cu atoms forming a dimer occupy an equivalent position. The orange color arrows represent the vibration directions of Cu atoms associated with the rattling mode to be described later. As shown in **Fig. 1c** and **Fig. 1d**, the intra-dimer distance is 2.48 Å, while the inter-dimer distances are 3.85 Å, 4.81 Å and 7.53 Å along *a*, *b* and *c* directions, respectively. In consideration of the isolated nature of the Cu dimer, the vibration mode of the dimer as a whole unit can be treated as a localized phonon. Later we will show that these localized phonons play a crucial role in suppressing the thermal conduction of $CuP_2$.

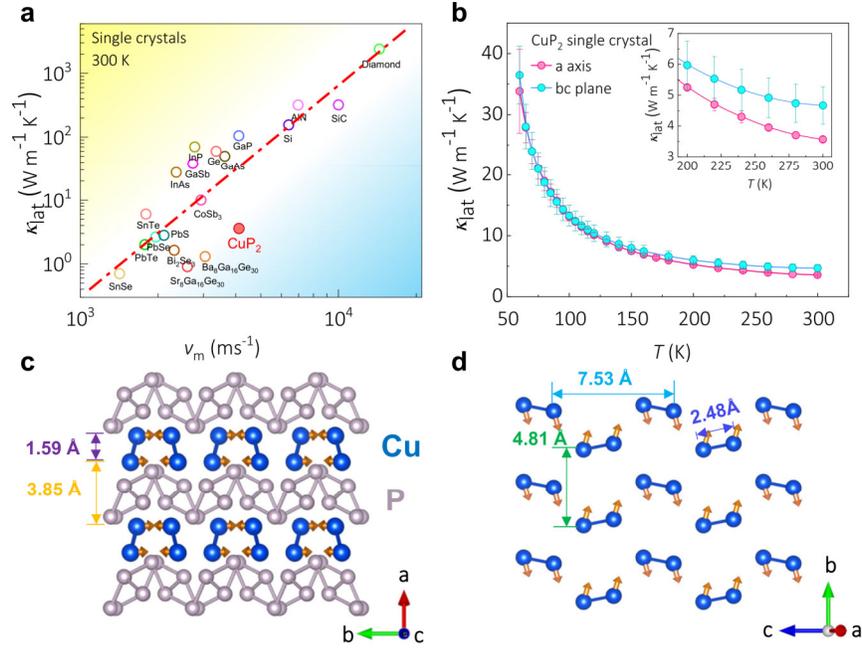

**Fig. 1 Lattice thermal conductivity and crystal structure of CuP$_2$. a.** A survey of lattice thermal conductivity of disorder-free materials versus mean sound speeds $v_m$. The detailed data are listed in Supplementary Table 1 (for polycrystalline samples, please refer to Supplementary Fig. 1 and Table 2). **b.** Temperature dependence of the lattice thermal conductivity of the CuP$_2$ single crystal is shown from 50 K to 300 K measured by a steady-state comparative method. The inset shows the data near room temperature. **c.** The layered structure with Cu dimer layers and P network layers are highlighted. The arrows in orange on Cu atoms represent the vibrational directions at the $\Gamma$ point of the optical phonon mode located at about 11 meV (see **Fig. 4**). The atomic motions of this mode is also displayed in Supplementary Video 1. **d**. The isolated Cu dimer layers with labelled intra- and inter-dimer distances.

Our neutron powder diffraction results confirm the crystal structure as previously reported[23]. Shown in **Fig. 2a** is the neutron powder diffraction pattern at room temperature (for data at lower temperatures, refer to Supplementary Fig. 2). The Rietveld refinement analysis suggests that the majority phase is the monoclinic CuP$_2$ while there is a minority phase of Cu$_3$P, whose volumetric fraction is only 2.79%. In addition, this powder sample, crushed from single crystals, is strongly textured along (100) because of the excellent ductility attributed to the layered lattice structure. It is important to notice that the background of the diffraction data is quite flat with no obvious diffuse scattering observed, which indicates no discernable disorder in the sample. The detailed structural parameters determined in the refinements are listed in Supplementary Table 3. The Debye-Waller factors, in particular, $U_{22}$ and $U_{33}$ of Cu atoms are much larger than those of P atoms. This indicates that there is a significant thermal motion of Cu atoms confined within the layers. The full temperature dependencies of the lattice parameters and Debye-Waller factors are summarized in Supplementary Fig. 3. In addition to the powder diffraction, our single crystal X-ray diffraction also confirms the structure. While the average structural study suggests the right phase and negligible disorder, we further employ the pair distribution function (PDF) analysis to directly confirm the disorder-free nature of the system. As shown in **Fig. 2b**, the X-ray structure factor $S^X(Q)$ is well normalized to 1 at high $Q$ at 300 K. Reduced PDF, $G^X(r)$, is subsequently obtained via Fourier transform of $S^X(Q)$. The inset of **Fig. 2b** presents the experimental $G^X(r)$ that is well reproduced by the $P2_1/c$ crystal model, verifying the disorder-free nature.

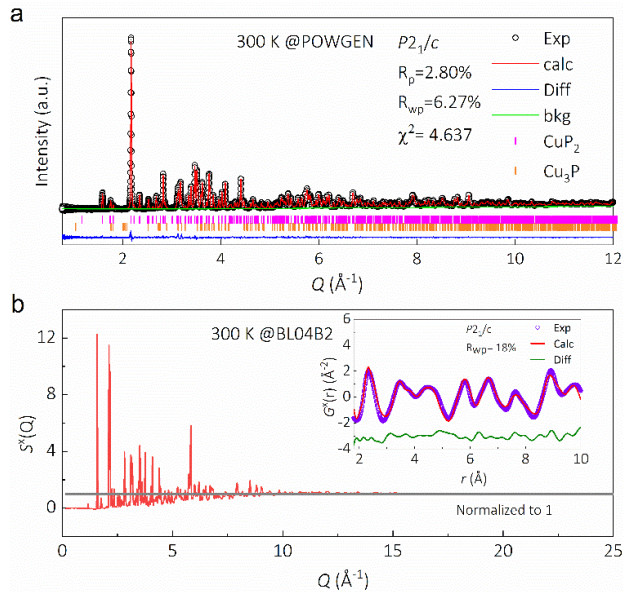

**Fig. 2 Average and local structures of CuP$_2$. a**. Neutron powder diffraction pattern and Rietveld refinement analysis at 300 K. **b**. The structure factor, $S^X(Q)$, obtained in X-ray total scattering at 300 K. The inset is the reduced PDF, $G^X(r)$, as well as the real-space refinement based on the $P2_1/c$ crystal model.

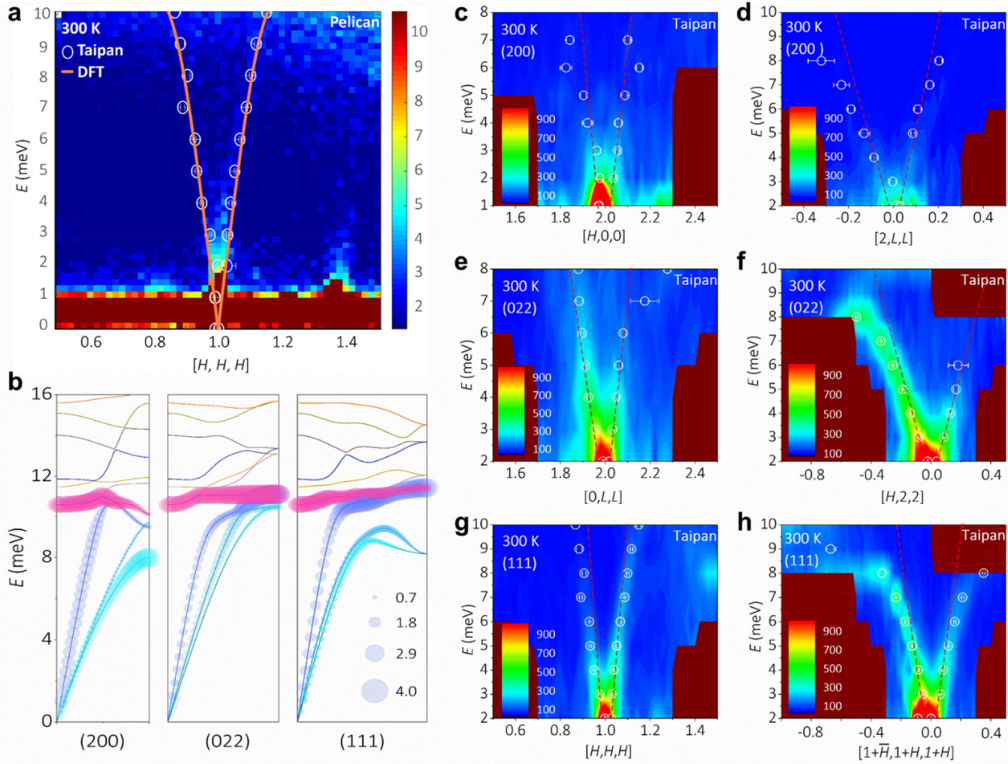

**Fig. 3 Phonon dispersions of CuP$_2$. a.** The phonon dispersion obtained at the time-of-flight neutron spectrometer-Pelican along [$H,H,H$] direction of Brillouin zone **G** = (111). The circles are the corresponding results from the triple-axis spectrometer-Taipan. The orange lines are the calculated dispersions. **b**. Calculated phonon dispersions of Brillouin zones **G** = (200), **G** = (022) and **G** = (111). The size of the shadow bubbles represents the magnitude of Grüneisen parameter ($\gamma$) of the related branches. **c-h.** Phonon dispersions of **G** = (200), **G** = (022) and **G** = (111) Brillouin zones collected on Taipan. The circles are the phonon energies determined by the spectral fitting to a Lorentzian function. The red dashed lines represent the linear fitting of phonon braches approaching the Brillouin zone centers to determine the sound speeds.

The above structural study excludes the disorder scattering as an origin of the low thermal conductivity. Now, we move to the lattice dynamics study for anharmonicity. We begin with a survey over a large reciprocal space of the dynamic structure function $S(\mathbf{Q}, E)$ versus momentum transfer (**Q**) and energy transfer ($E$) using the time-of-flight neutron spectrometer-Pelican. The measurements are carried out in the scattering plane defined by [*H*,0,0] and [0,*L*,*L*]. Shown in **Fig. 3a** is the phonon dispersion along [*H*,*H*,*H*] direction of the Brillouin zone **G** = (111). The branches of acoustic phonons are clearly observed, in agreement with the results from the triple-axis spectrometer-Taipan and the DFT calculations. Unfortunately, the optical phonon modes above 10 meV cannot be distinguished due to limited accessible energy range and energy resolution. For more details of dispersions of different Brillouin zones, please refer to Supplementary Figs. 4. Before we rationally perform very detailed INS measurements in a wider reciprocal space, the lattice dynamics are fully exploited using density-functional-theory (DFT) calculations. The DFT calculated phonons dispersions of Brillouin zones **G** = (200), **G** = (022), and **G** = (111) are plotted in **Fig. 3b** up to 16 meV while the complete dispersions are plotted in Supplementary Fig. 5. We can see the steep acoustic phonon branches that originate from the individual $\Gamma$ point and reach the zone boundaries at about 10 meV. Above the acoustic phonon branches, there are a few flat optical phonon bands up to 16 meV.

Then, we focus on the detailed investigations by conducting constant-*E* as well as constant-**Q** scans for Brillouin zones of **G** = (200), **G** = (022) and **G** = (111) using the thermal neutron triple-axis spectrometer-Taipan, in the same scattering plane as on Pelican. The scan directions are summarized in Supplementary Fig. 6. **Fig. 3c** and **d** show the dispersions along [*H*,0,0] and [0,*L*,*L*] directions of Brillouin zone **G** = (200) at 300 K obtained by constant-*E* scans, respectively. Note that while **Fig. 3c** represents the longitudinal component of the **G** = (200) Brillouin zone, some contributions of the longitudinal phonons are also included in **Fig. 3d** in addition to the major transverse components because of the reciprocal lattice setting in the monoclinic structure (see Supplementary Fig.6). The phonon intensity decays quickly with departure from the Brillouin zone center. To accurately determine the dispersion relationships, the phonon energies are determined by a spectral fitting to a Lorentzian function. The obtained peak positions are plotted on the contour plots as circles with error bars. Then, we fit the phonon energies to a linear function approaching the zone centers to obtain the experimental sound speeds (see the dash lines). The derived values are listed in Supplementary Table 3. It can be seen that the dispersion along [*H*,0,0] direction is much steeper than that along [0,*L*,*L*] direction, indicating larger sound speeds. Similar procedure is applied to the cases of **G** = (022) and **G** = (111), as shown in **Fig. 3e - f**, respectively. It can also be found that the phonon dispersions obtained from these two INS measurements and the theoretical calculated results are highly matched.

The temperature dependent lattice dynamics are considered first on the phonon density of state (PDOS) measurements on a powder sample in a wide temperature region. Shown in **Fig. 4a** are the PDOS at 200, 300, 400, 500 and 600 K, respectively. As temperature rises, it is clear that the whole profile is significantly broadened and several peaks are remarkably softened as an indication of strong anharmonicity. Compared with the DFT calculated PDOS, as plotted in **Fig. 4b**, it is identified that the Cu atoms mainly participate in the low-energy modes below 15 meV, while the high-energy optical modes are dominated by P atoms. This might be related to the fact that the covalent interactions of P networks are much stronger, in addition to the lighter atomic mass of P than that of the Cu dimer. Of particular interest is the experimental observed mode located at about 11 meV. Compared with the DFT calculations of the phonon dispersion (**Fig. 3b**) and PDOS (**Fig. 4b**), this mode is associated with Cu and has a flat dispersion, thus it must be a rattling mode from the Cu dimer (justified later). The strong anharmonicity of this rattling mode is evidenced by the large softening from 11.40 meV at 200 K to 10.67 meV at 600 K. Since the acoustic phonons and the nearby low frequency optical phonons are the main contributors to the thermal conductivity, we further investigate the anharmonic property of this Cu rattling mode in the following sections.

Focusing on the significantly softened mode at about 11 meV, detailed constant-**Q** scans were performed at 100, 300 and 450 K, respectively. **Fig. 4c** shows the dispersion along [*H*,0,0] direction of **G** = (022) at 100 K. Just above the intense acoustic phonon branch, a very flat optical phonon branch originates from the Brillouin zone center at about 12 meV to the Brillouin zone boundary at about 10 meV. This is the rattling mode of the Cu dimer. At higher energies, three more flat branches are further observed. The temperature dependencies of these modes are well tracked at the constant-**Q** scans at **q** = 0, 0.2, 0.3, as shown in **Fig. 4d – f**. Data at **q** = 0 clearly indicate a distinct behavior for the three optical modes identified. The energy of the first mode (rattling mode) is reduced from 11.16 meV at 100 K to 10.24 meV at 450 K, in a similar rate of 0.002 meV/K as observed with the PDOS results. In contrast, the other two modes at 14.53 and 15.53 meV at 100 K hardly shift with increasing temperature. At **q** = 0.2 and 0.3 shown in **Fig. 4e** and **f**, it is further noticed that the acoustic mode around 6 meV shows no noticeable shift in energy upon temperature increase in contrast to the rattling mode. This behavior is also clearly shown on more data as presented in Supplementary Figs. 7 – 9 for other Brillouin zones.

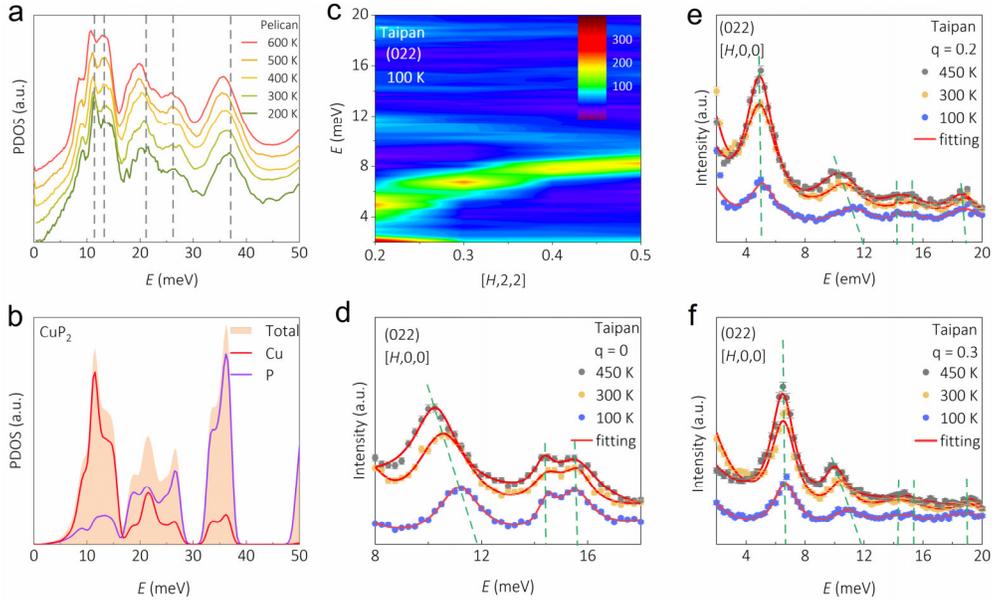

**Fig. 4 Highly anharmonic optical mode at 11 meV. a**. The experimental PDOS. **b.** Calculated total and partial PDOS. **c.** Phonon dispersion of **G** = (022) Brillouin zone along [$H$,0,0] at 100 K. **d-f**. Temperature dependence of phonon spectra for constant-**Q** scans of **G** = (022) Brillouin zone along [$H$,0,0] with **q** = 0, 0.2, 0.3 at 100, 300 and 450 K, respectively. The green dash lines highlight the energy shift of the peaks at heating.

**Table 1 Third-order Debye-Waller factors of Cu and P atoms determined from refinements of single crystal X-ray diffraction data.**

| $U$ (Å³) | Cu | P | |
| --- | --- | --- | --- |
| | Cu1 | P1 | P2 |
| $U_{111}$ | 0.0006(4) | 0.0001(7) | 0.0009(13) |
| $U_{222}$ | 0.0025(6) | 0.0011(9) | 0.00021(20) |
| $U_{333}$ | 0.00090(13) | 0.00021(20) | -0.0001(2) |

Apart from the remarkable temperature-induced softening of the mode as discussed above, the strong anharmonicity is also manifested by the huge Grüneisen parameter ($\gamma$)[24,25] calculated by DFT. In Fig. **3b**, the size of the shadow bubbles represents the magnitude of $\gamma$ with respect to a specific mode. Both the rattling mode (pink color) and the acoustic modes have very large $\gamma$. The largest $\gamma$ for the rattling mode is about 3.3 at **q** = 0.5, corresponding to the **G** = (022) Brillouin zone boundary, while $\gamma$ is almost larger than 2 in all reciprocal space covered. In particular, $\gamma$ is strikingly large for the LA branches (blue color) near the zone boundaries, where they tend to overlap with the rattling mode. Moreover, our single crystal X-ray diffraction measurement also suggests that Cu atoms have substantial third-order Debye-Waller factors, which provides further evidence of the strong anharmonicity of Cu-involved vibrations, as listed in **Table 1**. The values of $U_{222}$ and $U_{333}$ are 0.0025(6) Å³ and 0.00090(13) Å³ while $U_{111}$ is as small as 0.0006(4) Å³. Such a difference is consistent with the second-order Debye-Waller factors listed in Supplementary Fig. 2d. These noticeable third-order Debye-Waller factors indicate that the Cu atoms are weakly bonded in the *bc* plane with very restricted motions along *a* direction due to geometry confinement. As compared to Cu, the third-order Debye-Waller factors of P atoms are negligible within the error bars.

Henceforward, we discuss the nature of the highly anharmonic rattling mode at 11 meV and its impact on the system. Atomic motions of this mode are plotted at the $\Gamma$ point, as shown in **Fig. 1c** and **d**, which involve the vibrations of Cu dimers in between the P layers (see Supplementary Video 1). The justification of calling the vibration of the Cu dimers a rattling mode is further articulated here. Firstly, the Cu dimers behaving as rattlers have very weak bonding to the other dimers in the same layer. This is evidenced by the significant third-order Debye-Waller factors of Cu atoms that represent large-amplitude anharmonic vibrations in *b-c* plan, in agreement with the well-known skutterudites and clathrates compounds[13, 26, 27]. Secondly, the dispersion of this mode is quite flat as the group velocity is smaller than 10 m s$^{-1}$ showing an uncorrelated and localized feature. This is in close analogy to the ideal rattling scenario, appearing as a dispersionless Einstein mode with a constant vibration frequency. Thirdly, the anti-crossing or avoided-crossing of the acoustic mode with the rattling mode is also theoretically predicted in the compound, though not experimentally observed due to experiment limitations, as highlighted in Supplementary Fig. 5. The theory predicted 5 anti-crossing points located at **q** = 0.27 of **G** = (110), **q** = 0.32 of **G** = (100), **q** = 0.37 of **G** = (00½), **q** = 0.13 of **G** = ($\overline{½}$½0) and **q** = 0.32 of **G** = ($\overline{½}$00). This behavior also gives rise to the decrease of the group velocity of the acoustic mode around the zone boundaries. This rattling mode differs from the previous ones reported in skutterudites and clathrates systems. These compounds have typical cage-like frames with the heavy atoms filled into the lattice void, whereas CuP$_2$ crystallizes in a layered structure with Cu dimer as the rattler[28]. To distinguish with the conventional rattling mode, we term the behavior of CuP$_2$ as dimer rattling.

This rattling mode must play a dominating role in the thermal transport. We have systematically determined the sound speeds both experimentally and theoretically, as summarized in Supplementary Table 4. For example, $v_L = 6243$ m s$^{-1}$ and $v_T = 3192$ m s$^{-1}$ obtained from DFT calculation for the longitudinal and transverse acoustic branches of **G** = (200), respectively. This $v_L$ value is in excellent agreement with that determined by both fitting the INS experimental phonon dispersions and calculating from the Brillouin light scattering method (Supplementary Fig. 10), which are 6243 m s$^{-1}$ and 6275 m s$^{-1}$, respectively. The mean sound speed $v_m$ = 4115 m s$^{-1}$ is estimated using the equation $3v_m^{-3} = v_L^{-3} + 2v_T^{-3}$ (ref. 16) based on the theoretical results. As shown in **Fig. 1d**, unlike the ordinary materials that are distributed around the dash line, $CuP_2$ stands quite exceptional with the thermal conductivity one order of magnitude lower than GaAs having similar mean sound speed as $CuP_2$[5,6].

In summary, we have discovered the low thermal conducting property of $CuP_2$ and established a profound understanding on the fundamental physical mechanism by a comprehensive study on the atomic structures and lattice dynamics through neutron, X-ray scattering techniques and complementary DFT calculations. This system is manifested to be very anharmonic. The Cu atoms participate in a dimer rattling mode, which strongly scatters the LA phonons and leads to anti-crossing phenomena in the dispersion relationships. It is this mode that dominates the low thermal transport, counteracting the contribution of large mean sound speeds. The observed dimer rattling behavior in the open layered structures might represent emerging opportunity to rationally tailor thermal transport property of solids. The combined excellent acoustic and thermal insulation properties may find $CuP_2$ a promising material in some novel applications requiring both excellent mechanical sound transmission and heat insulation.

## Acknowledgements

This work was supported by the National Natural Science Foundation of China (Grant Nos. 11934007 and 11804346), the Key Research Program of Frontier Sciences, Chinese Academy of Sciences (Grant No. ZDBS-LY- JSC002), and the Liaoning

Revitalization Talents Program (Grant No. XLYC1807122). J.S.Z. was supported by an NSF grant (MRSEC DMR-1720595). We acknowledge beam time awarded by ANSTO (Proposal no. P7373), ORNL (proposal no. IPTS-21435.1), and SPring-8 (proposal no. 2019A1249). A portion of this research used resources at Spallation Neutron Source, a DOE Office of Science User Facility operated by the Oak Ridge National Laboratory. We thank Dr. Richard Mole for the help on the onsite data reduction and crystal alignment at Pelican as well as Dr. Guochu Deng for pre-aligning the crystal at the Joey Neutron Laue Camera.

## Contributions

B.L. and Zhidong Zhang conceived the idea. J.Q., Zongteng Zhang and W.R. synthesized the single crystals. X.C. and J.Z. measured the thermal conductivity of single crystals while L.F. and J.H. measured thermal conductivity of the polycrystalline pellet. Q.Z. collected the neutron powder diffraction data. X.J. collected the single crystal X-ray diffraction data and did the structural refinements. Y.C., K.O., S.H. and O.S. performed the synchrotron X-ray scattering and PDF analysis. Zhe Zhang, Zhao Zhang, B.L., D.Y., and S. D. conducted the INS measurements. G.C. performed the Brillouin light scattering experiments and determined the velocities. B.D. and T.Y. performed the DFT calculations. J.Q. analyzed the data and wrote the manuscript with B.L. and D.Y.

## Methods

**Sample preparation.** $CuP_2$ single crystals were grown by means of flux method[29]. Starting materials of Cu (purity: 99.999%), P (purity: 99.999%) and Sn (purity: 99.999%) in a molar ratio of 1:1:3 were placed in an alumina crucible and then sealed into an evacuated quartz tube. The mixture was placed into a box furnace and heated at 1233 K for 6 h. Then, it was cooled down to 873 K at a rate of 3 K/h. The excessive Cu and Sn eutectic flux was decanted in a centrifuge at 873 K. Single crystals were mechanically cleaved from the ingots. Crushed crystals were prepared for neutron and X-ray powder diffraction as well as for thermal conductivity measurement where a carbon-coated pellet ($\phi = 10$ mm) was used.

**Thermal conductivity measurement.** The total thermal conductivity ($\kappa_{tot}$) of the polycrystalline sample was obtained using the formula $\kappa_{tot} = DC_p d$, where $D$ is the thermal diffusion coefficient measured using the laser flash method (LFA457, NETZSCH, Germany), $C_p$ is the Dulong-Petit specific heat capacity and $d$ is the density calculated from the geometrical dimensions and mass. Thermal conductivity of the single crystals was measured in the temperature range from about 50 K to 300 K by a steady-state comparative method[30]. The bar-shaped CuP$_2$ samples with a dimension of about $0.5 \times 0.5 \times 3$ mm$^3$ were cut from as-grown crystals. The reference was a rod of constantan alloy with a diameter of 0.5 mm. The differential thermocouple was made of copper and constantan wires. The uncertainty of the thermal conductivity measurement is about 15%. The total thermal conductivity is expressed as a sum of lattice contribution ($\kappa_{lat}$) and electronic contribution ($\kappa_{el}$) for this compound:

$$\kappa_{tot} = \kappa_{lat} + \kappa_{el} \tag{1}$$

The electronic part $\kappa_{el}$ is proportional to the electrical conductivity $\sigma$ through the Wiedemann-Franz relation:

$$\kappa_{el} = L\sigma T \tag{2}$$

Where $L$ is Lorenz number[31]. The lattice thermal conductivity $\kappa_{lat}$ can be estimated by subtracting the electrical contribution $\kappa_{el}$. The electrical conductivity of CuP$_2$ is only about 8.09 $\Omega^{-1}$cm$^{-1}$ at 300 K[32]. $\kappa_{el}$ of CuP$_2$ is about 0.0059 W m$^{-1}$ K$^{-1}$ by assuming $L = 2.45$ W $\Omega$ K$^{-2}$ (ref. 33). In general, the true Lorenz number $L$ for most thermoelectric materials is lower than this value so that the electronic contribution to the thermal conductivity is negligible in this system[34].

**Neutron powder diffraction.** The neutron powder diffraction was performed at the time-of-flight powder diffractometer (POWGEN) at the Spallation Neutron Source (SNS) of Oak Ridge National Laboratory, USA[35]. The powder sample with the mass around 3 g was loaded into a vanadium container of 8mm diameter and measured in a Powgen Automatic Changer (PAC) covering the temperature region of 10 – 300 K. The data were collected with neutrons of central wavelength 0.8 Å. Constant temperature scans were conducted at 10, 50, 100, 200 and 300 K, respectively. GSAS[36] was used to refine the neutron powder diffraction patterns and the results are summarized in Fig. 2a, Supplementary Figs 2 and 3 as well as Table 3.

**Synchrotron X-ray powder scattering.** The high-energy X-ray powder diffraction experiment was carried out at the Beamline BL04B2 of SPring-8, Japan[37]. The $CuP_2$ powder was put into a quartz capillary ($\phi$ = 1 mm) and then fixed into the sample holder. An empty capillary was measured as a background. X-ray energy was fixed at 61.4 keV with a Si (220) monochromator. The energy resolutions $\Delta E/E$ was approximately $5\times10^{-3}$. A vacuum chamber with a Kapton window was used for minimizing the scattering background. A set of six point detectors were arranged horizontally to obtain $2\theta$ value up to 49° ($Q$ range up to 25 Å$^{-1}$). The real-space refinement was performed using PDFgui[38].

**Single crystal X-ray diffraction analysis.** The single crystal X-ray diffraction data was collected at a Pilatus CCD diffractometer equipped with graphite-monochromated Mo−Kα radiation (λ = 0.71073 Å) at 293 K. The crystal structure of $CuP_2$ was solved, and three-order atomic thermal displacement factors of all atoms were refined through full-matrix least-square technique on $F^2$. All of the calculations were performed using Jana2006 [39].

**INS measurements**. INS experiments were first performed on the time-of-flight cold-neutron spectrometer-Pelican[40,41], at Australian Nuclear Science and Technology Organisation (ANSTO), Australia. For the phonon density of states (PDOS) measurements, a powder sample was mounted in an annular aluminum sample cans with 1 mm gap. The sample can was attached to the cold head of a closed-cycle refrigerator which is capable of achieving sample temperature from 1.5 K to 800 K. The instrument was aligned for 4.69 Å (3.7 meV) incident neutrons. The resolution at the elastic line was 135 μeV. An empty can was subtracted to determine the background contribution and the data was normalized to a vanadium standard that had the same geometry as the sample can. All data manipulations were performed using Large Array Manipulation Program (LAMP)[42]. The scattering function $S(\mathbf{Q},\omega)$, as a function of scattering wave vectors ($\mathbf{Q}$) and phonon frequency ($\omega$), were measured on energy gain mode over a wide temperature range and then transformed to a generalized phonon density of states using formula (3), where $k_B$ is Boltzmann's constant, $T$ is temperature and $\hbar$ is the reduced Planck's constant

$$g(\omega) = \int \frac{\omega}{Q^2} S(Q,\omega)(1-e^{\frac{-\hbar\omega}{k_B T}}) dQ \qquad (3)$$

For phonon dispersion measurements, a single crystal weighted 0.7 gram was used and pre-orientated such that the scattering plane was defined by [$H$,0,0] and [0,$L$,$L$]. A large reciprocal space was covered by rotating the sample over 100 degrees in a step of 1 degree around the axis perpendicular to the scattering plane. The incident neutron wavelength is 2.345 Å (14.8 meV), corresponding to the second order reflection of the Pelican HOPG monochromators configured at 4.69 Å. The sample temperature was set at 300 K. The data were reduced by LAMP and the whole $S(\mathbf{Q},\omega)$ was generated using HORACE[43], which was also used to visualize the phonon dispersions. After the general survey over many Brillouin zones using Pelican instrument, we focused on several selected Brillouin zones $\mathbf{G}$ = (200), $\mathbf{G}$ = (022) and $\mathbf{G}$ = (111) on the thermal neutron triple-axis spectrometer-Taipan at ANSTO[44,45]. The energies of the incident and scattered neutrons were defined by a double-focused PG (002) monochromator and analyser. We used the open geometry of the neutron beam with a virtual source width of 10 mm at the beam exit from the reactor and a slit of 20 mm at the detector. The measurements were performed with a fixed final neutron energy of 14.87 meV. After the sample, a highly oriented pyrolytic graphite filter (HOPG) was placed to remove higher-order reflections from the scattered beam. The sample was aligned in the same scattering plane as in the measurements at Pelican. The phonons were measured along [$H$,0,0], [$H$,$H$,$H$] and [0,$L$,$L$] directions. A standard cryofurnace was used to access the temperatures of 100, 300 and 450 K. The experimental data were fitted to a Lorentzian function using PAN of DAVE[46].

**Brillouin light scattering.** Brillouin light scattering measurements were performed at room temperature using a Sandercock-type six-pass tandem Fabre-Perot (TFP-2) as a spectrometer and a 532 nm laser as a light source. It is a well-established technique for measuring sound velocities[47]. In backscattering geometry, longitudinal acoustic mode sound velocity ($v_L$) can be calculated as

$$v_L = \frac{f_B \lambda_0}{2n} \tag{4}$$

From the measured Brillouin frequency shift $f_B$, using this equation, the refractive index $n$ = 4.04 and the laser wavelength $\lambda_0$ = 532 nm, the longitudinal acoustic-wave velocity was determined to be $v_L$ = 6275 m s$^{-1}$.

**DFT calculations.** All the calculations in this work were performed using the *ab initio* density functional theory as implemented in the VASP code[48]. The projector augmented wave (PAW) pseudopotentials[49] and the Perdew-Burke-Ernzerhof (PBE)[50] functional within the General Gradient Approximation (GGA) were used to take care of electron-ion and inter-electron exchange-correlation interactions, respectively. The wave functions were expanded using plane waves with energy cutoff of 500 eV, and the electronic energy convergence was set to be 10$^{-8}$ eV. The Brillouin zone of the primitive unit cell was sampled in the Γ-centered 8×10×15 k-point mesh for structural optimization until all the atomic force is less than 0.001 eV/Å. The phonon dispersion relation was calculated using Phonopy package[51] combining with VASP using 2×2×2 supercell with 96 atoms. The Γ-centered 2×2×2 q-point mesh was used.

## Supplementary information

Supplementary Figures 1- 10

Supplementary Tables 1-4

Supplementary Videos 1- 3